\documentclass[epjST]{svjour}
\usepackage{graphics}
\usepackage{amsmath}
\usepackage{amssymb}
\usepackage{mathtext}
\usepackage{color}
\begin{document}

\title{Specific interface area and self-stirring in a two-liquid system experiencing intense interfacial boiling below the bulk boiling temperatures of both components}
%\date{\today}
\author{Denis S.\ Goldobin\inst{1}\fnmsep\inst{2}\fnmsep\thanks{\email{Denis.Goldobin@gmail.com}}\and
        Anastasiya V.\ Pimenova\inst{1}\fnmsep\thanks{\email{Anastasiya.Pimenova@gmail.com}}}
\institute{Institute of Continuous Media Mechanics, UB RAS,
             Perm 614013, Russia\and
           Department of Theoretical Physics,
             Perm State University, Perm 614990, Russia
           }

\abstract{
We present an approach to theoretical assessment of the mean specific interface area $(\delta{S}/\delta{V})$ for a well-stirred system of two immiscible liquids experiencing interfacial boiling. The assessment is based on the balance of transformations of mechanical energy and the laws of the momentum and heat transfer in the turbulent boundary layer. The theory yields relations between the specific interface area and the characteristics of the system state. In particular, this allows us to derive the equations of self-cooling dynamics of the system in the absence of external heat supply.
The results provide possibility for constructing a self-contained mathematical description of the process of interfacial boiling. In this study, we assume the volume fractions of two components to be similar as well as the values of their kinematic viscosity and molecular heat diffusivity.
}

%\begin{keyword}
%Interfacial boiling \sep Two-liquid systems \sep Specific interface area
%\end{keyword}

\maketitle

\section{Introduction}
For a well-stirred multiphase fluid systems the mean interface area per unit volume, {\it or} the specific interface area, say $S_{\!V}\equiv(\delta{S}/\delta{V})$, is a significant characteristic of the state. It becomes even more important for the systems where the interface is active chemically or physically (for instance, interface of phase transition). The systems of immiscible liquids experiencing interfacial boiling~\cite{Krell-1982,Geankoplis-2003,Simpson-etal-1974,Celata-1995,Roesle-Kulacki-2012-1,Roesle-Kulacki-2012-2,Sideman-Isenberg-1967,Kendoush-2004,Filipczak-etal-2011,Gordon-etal-1961,Prakash-Pinder-1967-1,Prakash-Pinder-1967-2,Pimenova-Goldobin-JETP-2014,Pimenova-Goldobin-2014-2} are an example of the systems where parameter $S_{\!V}$ has essential control on the evolving process.

It is well known that the boiling at the interface between two immiscible liquids can occur bellow the bulk boiling temperatures of both components ({\it e.g.}, see~\cite{Krell-1982,Geankoplis-2003}). This phenomenon is widely reported in literature and finds many applications in industry. Species from both liquid phases evaporate into the vapour layer forming between them. As a result, the condition for the vapour layer growth ({\it e.g.}, see~\cite{Krell-1982}) is
\[
n_1^{(0)}(T)+n_2^{(0)}(T)\ge\frac{P_0}{k_\mathrm{B}T}\,,
\]
where $n_j^{(0)}(T)$ is the number density of the saturated vapour of specie $j$ at absolute temperature $T$, $P_0$ is the atmospheric pressure, and $k_\mathrm{B}$ is the Boltzmann constant. The minimal interfacial boiling temperature $T_\ast$ corresponds to equality in the latter condition. Simultaneously, the bulk boiling of liquid component $j$ requires $n_j^{(0)}(T)$ alone to exceed $P_0/(k_\mathrm{B}T)$, meaning that interfacial boiling starts below the bulk boiling points of both components. The interfacial boiling process is essentially controlled by the specific interface area $S_{\!V}$.

The problem of calculation of $S_{\!V}$ cannot be addressed rigourously. Moreover, any direct numerical simulation, being extremely challenging and CPU-time consuming, will provide results pertaining to specific system set-ups. Some general assessments on $S_{\!V}$ can be highly beneficial. In this paper we perform these assessments for the process of direct contact boiling in a system of two immiscible liquids.

The big picture of mechanical processes in the system is as follows. At the direct contact interface, a vapour layer grows and produces bubbles which breakaway from the interface and rise. The presence of vapour bubbles changes the fluid buoyancy and performs a ``stirring'' of the system. This stirring enforces increase of the contact area $S$, while surface tension and gravitational segregation of two liquids tend to minimize this contact area.

For a two-liquid system experiencing direct contact boiling, the quantity of our interest depends on parameters of liquids and characteristics of the evaporation process, which are controlled by the mean temperature excess above $T_\ast$ and the bubble production rate~\cite{Filipczak-etal-2011}. In our study, we assume the volumes of both components to be commensurable; no phase can be considered as a medium hosting dilute inclusions of the other phase. Hence, the characteristic width of the neighborhood of the vapour layer, beyond which the neighborhood of another vapour layer sheet lies, is
\begin{equation}
\nonumber
H\sim 1/S_{\!V}\,.
\end{equation}
%(see Fig.~\ref{fig2}).
In this paper, as a first approach to the problem, we simplify our consideration, restricting ourselves to the case of liquids with similar values of those physical parameters, which control the properties of the turbulent boundary layer and the heat transfer in it: kinematic viscosity and molecular heat diffusivity. The assumptions of similar volumes of both fractions and similar physical parameters make our problem symmetric with respect to interchange of two liquids and significantly simplify the study.
%The relationship between $H_1$ and $H_2$ is
%$$
%\frac{H_1}{H_2}=\frac{\phi_1}{\phi_2}=\frac{\phi_1}{1-\phi_1}\,,
%$$
%where $\phi_j$ is the volumetric fraction of the $j$-th liquid in
%the system. It will be convenient to use
%\begin{equation}
%H_j\sim\phi_j\left(\frac{\delta{S}}{\delta{V}}\right)^{-1}.
%\nonumber
%\end{equation}

The process of boiling of a mixture above the bulk boiling temperature of the more volatile liquid is well-addressed in the literature~\cite{Simpson-etal-1974,Celata-1995,Roesle-Kulacki-2012-1,Roesle-Kulacki-2012-2,Sideman-Isenberg-1967,Kendoush-2004,Filipczak-etal-2011}. Hydrodynamic aspects of the process of boiling below the bulk boiling temperature have been theoretically studied in~\cite{Pimenova-Goldobin-JETP-2014,Pimenova-Goldobin-2014-2,Pimenova-Goldobin-2016}. While in~\cite{Pimenova-Goldobin-JETP-2014,Pimenova-Goldobin-2014-2} the specific interface area $S_{\!V}$ was treated as a control parameter of the system dynamics, here we construct a theory allowing to find the relationships between this parameter and the macroscopic characteristics of the state of a well-stirred system. Specifically, in what follows, we perform an analytical assessment of the dependence of $S_{\!V}$ on the evaporation rate (or heat influx) and mean temperature and derive equations of self-cooling in the absence of external heat sources for a well-stirred system.

Our work heavily relies on the theory of turbulent boundary layer~\cite{Karman-1930,Prandtl-1933,Landau-Lifshitz,Schlichting-Gersten} which is employed for calculation of the rates of transfer of kinetic flow energy and heat from the bulk of phases towards the boiling interface.

%%%%%%%%%%%%%%%%%%%%%%%%%%%%%%%%%%%%%%%%%%%%%%%%%%%%%%%%%%%%%%%%%%%%%%%%%%%%%
%%%%%%%%%%%%%%%%%%%%%%%%%%%%%%%%%%%%%%%%%%%%%%%%%%%%%%%%%%%%%%%%%%%%%%%%%%%%%
\begin{figure}[t]
\center{
\resizebox{0.32\columnwidth}{!}{
\includegraphics%[width=0.25\textwidth]
{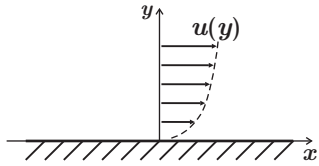}}}
\caption{Sketch of the turbulent boundary layer near the flat rigid wall and notations}
\label{fig1}
\end{figure}
%%%%%%%%%%%%%%%%%%%%%%%%%%%%%%%%%%%%%%%%%%%%%%%%%%%%%%%%%%%%%%%%%%%%%%%%%%%%%
%%%%%%%%%%%%%%%%%%%%%%%%%%%%%%%%%%%%%%%%%%%%%%%%%%%%%%%%%%%%%%%%%%%%%%%%%%%%%

\section{Turbulent boundary layer}
For liquids with viscosity properties similar to that of water, the viscous boundary layer of flows turns out to be very thin compared to the realistic scales of folds of the interface\footnote{This statement will be underpinned with estimates below in the text.}, meaning the flow in the liquid bulk is either inviscid (practically improbable) or turbulent (which is more plausible for the system under our consideration). Hence, one has to consider the turbulent currents near the interface; the description of properties of these currents is available from the theory of a turbulent boundary layer.

The theory of a turbulent boundary layer has been developed by Karman~\cite{Karman-1930} and Prandtl~\cite{Prandtl-1933}, and is also well presented in book~\cite{Landau-Lifshitz}, on which we heavily relay in this section.
For a turbulent boundary layer near a flat rigid wall beyond the `viscous sublayer'---a thin vicinity of the boundary where molecular and turbulent viscosities are of the same order of magnitude---the `macroscopic' (averaged over pulsations) flow $u$ is a shear one (tangential to the boundary) and is controlled by the fluid density $\rho$, momentum flux to the boundary $\varPi$, and the distance from the boundary $y$. Specifically,
\begin{equation}
\frac{du}{dy}=\frac{u_\ast}{\varkappa y}\,,
\label{eq101}
\end{equation}
where the reference turbulent pulsation velocity $u_\ast$ is introduced as follows: $\varPi=\rho u_\ast^2$, $\varkappa$ is the dimensionless Karman constant determined experimentally,
\[
\varkappa\approx 0.4\,.
\]

Generally,
\begin{equation}
u=u_\ast f(\xi)\,,\qquad \xi=yu_\ast/\nu\,,
\label{eq102}
\end{equation}
where $\nu$ is the kinematic viscosity. Beyond the viscous sublayer ({\it i.e.}, for $\xi\gtrsim1$), Eq.\,(\ref{eq101}) and experimental data yield
\begin{equation}
u=\frac{u_\ast}{\varkappa}\ln\frac{yu_\ast}{\xi_0\nu}\,,
\label{eq103}
\end{equation}
where constant $\xi_0\approx0.13$ is determined from experiments. The average energy flux to the boundary is
\begin{equation}
q_u=u\varPi.
\label{eq104}
\end{equation}

%%%%%%%%%%%%%%%%%%%%%%%%%%%%%%%%%%%%%%%%%%%%%%%%%%%%%%%%%%%%%%%%%%%%%%%%%%%%%
%%%%%%%%%%%%%%%%%%%%%%%%%%%%%%%%%%%%%%%%%%%%%%%%%%%%%%%%%%%%%%%%%%%%%%%%%%%%%
\begin{figure}[t]
\center{
\resizebox{0.33\columnwidth}{!}{
\includegraphics%[width=0.27\textwidth]
{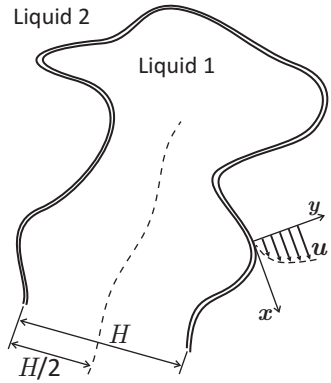}}}
\caption{Sketch of the vapour layer between two immiscible liquids}
\label{fig2}
\end{figure}
%%%%%%%%%%%%%%%%%%%%%%%%%%%%%%%%%%%%%%%%%%%%%%%%%%%%%%%%%%%%%%%%%%%%%%%%%%%%%
%%%%%%%%%%%%%%%%%%%%%%%%%%%%%%%%%%%%%%%%%%%%%%%%%%%%%%%%%%%%%%%%%%%%%%%%%%%%%

For the temperature field the average profile in the turbulent boundary layer is additionally controlled by the Prandtl number $\Pr=\nu/\chi$ and is affected by the fact the temperature field is a passive scalar, while the average velocity is a vector quantity, which is also correlated with the turbulent vortex cascade,
\begin{equation}
T=\beta\frac{q_T}{\varkappa\rho c_P u_\ast}\left[\ln\frac{yu_\ast}{\nu}+f_T(\Pr)\right]+T_0\,,
\label{eq105}
\end{equation}
where $T_0$ is the boundary temperature, $q_T$ is the heat energy flux towards the boundary, $c_P$ is the specific heat at constant pressure; constant $\beta=0.9$ and function $f_T(\Pr)$ (instead of $\ln(1/\xi_0)$, as compared to Eq.\,(\ref{eq103})) represent the mentioned differences between $u$ and $T$~\cite{Landau-Lifshitz,Schlichting-Gersten}.
% For reference, $f_T(0.7)=1.5$ (air) and $f_T(\Pr\gg1)\propto\Pr^{3/4}$ [Landau].

For realistic situations, $u_\ast$ is in range $0.1$--$0.2\,\mathrm{m/s}$ and the argument of the logarithm in Eq.\,(\ref{eq103}) is of the order of magnitude of 1 for $y\sim10^{-6}\,\mathrm{m}$. On this scale the free interface is inflexible because of the surface tension. Interface deformations become plausible for the scale of $1\,\mathrm{mm}$, where the logarithm argument $(yu_\ast)/(\xi_0\nu)\sim10^3$ and the further increase of the distance $y$ by a factor $10$ causes the relative change of the logarithm only by $1/4$. Hence, the dominant change of average fields of flow and temperature occurs on the scales, where the interface is nearly inflexible, and the theory of the turbulent boundary layer can be relevant for our system with the liquid--liquid interface in the role of the boundary (Fig.\,\ref{fig2}). However, one should keep in mind, that on the rigid wall the velocity is strictly zero, while for a free inflexible interface the vortexes with currents along the interface are admitted. Therefore, the quantitative characteristics of the boundary layer profiles can be altered compared to their values for the case of a rigid wall. Since constants $\varkappa$ and $\beta$ are related to the scale-independent properties of profiles, while the difference between the cases of a rigid wall and a free inflexible interface influences the processes within the viscous sublayer, one can expect the change of $\xi_0$ and $f_T(\Pr)$, but $\varkappa$ and $\beta$ must hold the same. The latter remark is important for us, because of unfortunate lack of knowledge on the turbulent boundary layer near a free inflexible surface.

Employing the knowledge on the turbulent boundary layer properties, one can estimate the rate of viscous dissipation of the kinetic energy of the liquid flow and the heat flux to the interface.

\section{Viscous energy dissipation and heat transfer in the turbulent boundary layer}
The kinetic energy flux from the bulk of certain liquid component towards the interface can be assessed on the basis of relation~(\ref{eq104}). On the other hand, this flux is a viscous loss of the kinetic energy of macroscopic liquid flow in the bulk;
\begin{equation}
q_u(y=H/2)\approx\frac{1}{2}\frac{\delta\dot{W}_\mathrm{visc}}{\delta S}\,,
\label{eq106}
\end{equation}
where $\delta\dot{W}_\mathrm{visc}$ is the viscous energy loss per area $\delta S$ of the interface. Hence,
\begin{eqnarray}
\frac{\delta\dot{W}_\mathrm{visc}}{\delta V}&\approx&
\left(\frac{\delta S}{\delta V}\right)\,2q_u(y\!=\!H/2)
\nonumber\\[5pt]
&=&2S_{\!V} \varPi u(y\!=\!H/2)
\nonumber\\[5pt]
&=&S_{\!V}\frac{2\rho u_\ast^3}{\varkappa}\ln\frac{Hu_\ast}{2\xi_0\nu}\,.
\label{eq107}
\end{eqnarray}

The space-average kinetic energy of liquid flow, say $W_\mathrm{liq,k}$, is mainly contributed by average flow $u$, as the contribution of turbulent pulsation flow can be neglected~\cite{Landau-Lifshitz}; therefore,
\begin{equation}
\frac{\delta W_\mathrm{liq,k}}{\delta V}\approx
\frac{2}{H}\int\limits_{0}^{H/2}\frac{\rho u^2}{2}\mathrm{d}y
\approx
\frac{\rho u_\ast^2}{2\varkappa^2}\ln^2\frac{Hu_\ast}{e\xi_0\nu}\,,
\label{eq108}
\end{equation}
where $e$ is the Euler's number.

One can evaluate the average temperature from Eq.\,(\ref{eq105}), finding the relation between $\langle{T}\rangle$ and $q_T$;
\begin{equation}
\langle{T}\rangle=\frac{2}{H}\int\limits_0^{H/2}T\mathrm{d}y\approx
T_0+\frac{\beta\,q_T}{\varkappa\rho c_P u_\ast}\left(
\ln\frac{Hu_\ast}{2\nu}+f_T(\Pr)-1\right).
\label{eq109}
\end{equation}
We assume the interface temperature to be equal to the minimal temperature required for the growth of the vapour layer, $T_0=T_\ast$. Indeed, with the data provided in~\cite{Pimenova-Goldobin-2014-2}, one can estimate the typical difference $(T_0-T_\ast)$ observed in experiments as $\sim10^{-3}\,\mathrm{K}$. Hence, we can neglect it for our consideration in this paper. 
From Eq.\,(\ref{eq109}), one can evaluate the dependence of heat flux on the average overheat ({\it or} the temperature excess above $T_\ast$) $\langle\varTheta\rangle=\langle{T-T_\ast}\rangle$;
\begin{equation}
q_T\approx\frac{\displaystyle\varkappa\rho c_P u_\ast \langle\varTheta\rangle}
 {\displaystyle\beta\left(\ln\frac{Hu_\ast}{2\nu}+f_T(\Pr)-1\right)}\,.
\label{eq110}
\end{equation}

\section{Energy flux balance in a well-stirred system}
Let us derive relationships between the macroscopic parameter $S_{\!V}=(\delta{S}/\delta{V})$, so-called specific interface area, of the system state and the heat influx rate per unit volume $\dot{Q}_V=\delta{Q}/(\delta{V}\delta{t})$ for a statistically stationary process of interfacial boiling. Statistical stationarity of the enforced stirring process implies as well balance of mechanical energy fluxes.

One should distinguish macroscopic degrees of freedom and internal (thermodynamic) degrees of freedom. It is important, because while the viscous dissipation of the macroscopic kinetic energy into the internal energy is possible, the direct transformation of the internal energy into the macroscopic one is forbidden by general principles of thermodynamics. The only mechanism of the energy supply for macroscopic motion is the buoyancy of generated vapour bubbles. Let us now compare the amount of energy in macroscopic and internal degrees of freedom.
The energy of thermal motion of atoms corresponds to characteristic atom velocities $10^2-10^3\,\mathrm{m/s}$. Particularly for water, the latent heat of evaporation is even significantly larger than the kinetic energy of thermal motion of its atoms at $T=300\,\mathrm{K}$.
For macroscopic degrees of freedom the mechanical motion is enforced by potential forces of gravity and surface tension. Hence, the net energy associated with these degrees of freedom is of the same order of magnitude as the macroscopic kinetic energy. For the realistic systems of our interest, the macroscopic stirring flow of velocity $1\,\mathrm{m/s}$ can be considered as very strong, which means that the energy of macroscopic flow is typically by a factor $10^5$ or more smaller than the internal energy.
Thus, the macroscopic degrees of freedom make vanishingly small contributions into the balance of internal energy.
Summarizing,
\\
(i)~the supply of the macroscopic mechanical energy into system is associated solely with the buoyancy of generated vapour bubbles,
\\
(ii)~the amount of generated vapour bubbles is determined by the heat transfer to the boiling interface.

The flow and consequent stirring in the system are enforced by the buoyancy of the vapour bubbles, while other mechanisms counteract the stirring of the system. These other mechanisms are gravitational stratification of two liquids, surface tension tending to minimise the interface area and viscous dissipation of the flow energy.
As discussed above, all the heat inflow into the system can be considered to be spent for the vapour generation;
 $\dot{Q}_VV\longrightarrow(\Lambda_1n_1^{(0)}+\Lambda_2n_2^{(0)})\dot{V}_\mathrm{vap}$,
where $V$ is the system volume, $\dot{V}_\mathrm{vap}$ is the volume of the vapour produced in the system per unit time, $\Lambda_j$ is the enthalpy of vaporization per one molecule of liquid $j$, and $n_j^{(0)}$ is the saturated vapour pressure of liquid $j$. Thus,
\begin{align}
\dot{V}_\mathrm{vap}=\frac{\dot{Q}_V\,V}
 {\Lambda_1n_1^{(0)}+\Lambda_2n_2^{(0)}}\,.
\label{eq111}
\end{align}

The potential energy of buoyancy of rising vapour bubbles $\rho V_\mathrm{vap}gh/2$ (where $h$ is the linear size of the system, $h\sim V^{1/3}$, $\rho$ is the average density of liquids, the vapour density is zero compared to the liquid density) is converted into the kinetic energy of liquid flow, the potential energy of a stirred state of the two-liquid system, the surface tension energy and dissipated by viscosity forces. In a statistically stationary state, the mechanical kinetic and potential energies do not change averagely over time and all the mechanical energy influx is to be dissipated by viscosity;
$$
\rho V_\mathrm{vap}gh/2\longrightarrow\dot{W}_\mathrm{visc}\tau\,,
$$
where $\dot{W}_\mathrm{visc}$ is the rate of viscous dissipation of energy, $\tau$ is the time of generation of the vapour volume $V_\mathrm{vap}$, $V_\mathrm{vap}=\dot{V}_\mathrm{vap}\tau$. Hence,
\begin{align}
\rho\dot{V}_\mathrm{vap}g\frac{h}{2}=\dot{W}_\mathrm{visc}\,.
\label{eq112}
\end{align}
In turn, the viscous dissipation of the kinetic energy of flow $\dot{W}_\mathrm{visc}$ is determined by Eq.\,(\ref{eq107}).

Further, we have to establish the relationship between the flow kinetic energy and the mechanical potential energy in the system. Rising vapour bubbles ``pump'' the mechanical energy into the system, while its stochastic dynamics is governed by interplay of its flow momentum and the forces of the gravity and the surface tension on the interface. Hamiltonian systems with huge number of degrees of freedom experience thermalization, or one can say, they tend to the state of thermodynamic equilibrium. In thermodynamic equilibrium, the total energy is strictly equally distributed between potential and kinetic energies related to quadratic terms in Hamiltonian. (The latter statement is frequently simplified to a less accurate statement, that energy is equally distributed between kinetic and potential energies associated with each degree of freedom.) Although in turbulent systems the viscosity plays a crucial role, the kinetic energy is dissipated only on the edges of wavenumber spectrum, meaning the system is weakly dissipative and its dynamics is nearly conservative. Being not exactly in the case where one can rigorously speak of thermalization of the stochastic Hamiltonian system dynamics, we still may assess the kinetic energy of flow to be of the same order of magnitude as the mechanical potential energy of the system. Thus,
\begin{align}
W_\mathrm{liq,k}\sim W_{\mathrm{liq,p}g}+W_{\mathrm{liq,p}\sigma}\,,
\label{eq113}
\end{align}
where $W_{\mathrm{liq,p}g}$ and $W_{\mathrm{liq,p}\sigma}$ are the gravitational potential energy and the surface tension energy, respectively. We set the zero levels of these potential energies at the stratified state of the system with a flat horizontal interface.

The gravitational potential energy of the well-stirred state with uniform distribution of two phases over height is
$$
W_{\mathrm{liq,p}g}=\Delta\rho Vg\frac{h}{2}\,,
$$
where $\Delta\rho$ is the component density difference. The surface tension energy is
$$
W_{\mathrm{liq,p}\sigma}\approx(\sigma_1+\sigma_2)S_{\!V}V\,,
$$
where we have neglected the interface area of the stratified state compared to the area $S_{\!V}V$ in a well-stirred state. Due to the presence of the vapour layer between liquids the effective surface tension coefficient of the interface is $(\sigma_1+\sigma_2)$ but not $\sigma_{12}$ as it would be in the absence of the vapour layer.

\section{Specific interface area $S_{\!V}$ and balance of energy transfer}
\subsection{$S_{\!V}$ as a function of average overheat $\langle\varTheta\rangle$}
From Eqs.\,(\ref{eq107}), (\ref{eq112}), and (\ref{eq111}), one can find for $\dot{Q}_V$:
\begin{equation}
\dot{Q}_V\approx\frac{\Lambda_1n_{1}^{(0)}+\Lambda_2n_{2}^{(0)}}{gh}
 S_{\!V}\frac{4u_\ast^3}{\varkappa}\ln\frac{Hu_\ast}{2\xi_0\nu}\,.
\label{eq114}
\end{equation}
On the other hand, the heat inflow to the interface area within the volume $\delta{V}$, $\dot{Q}_V\delta{V}$, is contributed by heat flux $q_T$ from two sides of the interface, $2q_T\delta{S}$, {\it i.e.}, $\dot{Q}_V=2S_{\!V}q_T$, and, with Eq.\,(\ref{eq110}), one can evaluate
\begin{equation}
\dot{Q}_V=S_{\!V}\frac{\displaystyle 2\varkappa\rho c_P u_\ast\langle\varTheta\rangle}
 {\displaystyle\beta\left(\ln\frac{Hu_\ast}{2\nu}+f_T(\Pr)-1\right)}
\approx S_{\!V}\frac{\displaystyle 2\varkappa\rho c_P u_\ast\langle\varTheta\rangle}
 {\displaystyle\beta\ln\frac{Hu_\ast}{2\xi_0\nu}}\,.
\label{eq115}
\end{equation}
Matching heat consumption (\ref{eq114}) for the vapour generation and turbulent heat transfer to the interface (\ref{eq115}), one obtains
\begin{eqnarray}
\langle\varTheta\rangle\approx\frac{\Lambda_1n_{1}^{(0)}+\Lambda_2n_{2}^{(0)}}{\rho c_Pgh}
2\beta\frac{u_\ast^2}{\varkappa^2}\ln^2\frac{Hu_\ast}{2\xi_0\nu}
\nonumber\\[5pt]
\approx\frac{4\beta(\Lambda_1n_{1}^{(0)}+\Lambda_2n_{2}^{(0)})}{\rho^2c_Pgh}
\frac{\delta W_\mathrm{liq,k}}{\delta V}\,.
\label{eq116}
\end{eqnarray}
Eq.~(\ref{eq113}) provides the relation between $W_\mathrm{liq,k}$ and potential energy, which is an explicit function of macroscopic parameter $S_{\!V}$. Hence,
\begin{equation}
\langle\varTheta\rangle\approx
\varTheta_g\left[1+\frac{2}{k_{12}^2h}S_{\!V}\right]\,,
\label{eq117}
\end{equation}
where
\[
\varTheta_g=\frac{2\beta(\Lambda_1n_{1}^{(0)}+\Lambda_2n_{2}^{(0)})\Delta\rho}{\rho^2c_P}
\]
and the characteristic wavenumber for interfacial waves
\[
k_{12}=\sqrt{\frac{(\rho_2-\rho_1)g}{\sigma_1+\sigma_2}}\,,
\]
where, as noted above, the effective surface tension coefficient of the contact interface in the presence of the vapour layer is $(\sigma_1+\sigma_2)$~\cite{Pimenova-Goldobin-2016}.
Noteworthy, the relative importance of the first and second terms in the brackets in Eq.~(\ref{eq117}) depends on the vertical size of the system $h\sim V^{1/3}$.

For the $n$-heptane--water system, $l_{k_{12}}\equiv1/k_{12}\approx0.5\,\mathrm{cm}$ and $\varTheta_g\approx0.346\,\mathrm{K}$~\cite{Pimenova-Goldobin-2014-2}. For a well-stirred system the distance between sheets of the folded interface $H\sim S_{\!V}^{-1}\ll h$. One can distinguish two limiting cases;
\\
(1)~$h/S_{\!V}\ll l_{k_{12}}^2$, which corresponds to the case of a surface-tension dominated system,
\\
(2)~$h/S_{\!V}\gg l_{k_{12}}^2$, which corresponds to the case of a gravity-driven system.
\\
The temperature $\varTheta_g$ is remarkably small compared to the maximal overheat $\varTheta$ of the $n$-heptane--water system $20\,\mathrm{K}$, which can be attained before the bulk boiling of components can occur. Thus, according to Eq.~(\ref{eq117}), the boiling regime will be typically surface-tension dominated (since the second term in the brackets will be typically large compared to $1$).

An inverse form of Eq.~(\ref{eq117}) provides $S_{\!V}$ as a function of the average overheat $\langle\varTheta\rangle$;
\begin{equation}
S_{\!V}\approx\frac{k_{12}^2h}{2}\left(\frac{\langle\varTheta\rangle}{\varTheta_g}-1\right).
\label{eq118}
\end{equation}

\subsection{Systems driven by heat inflow $\dot{Q}_V$}
Let us rewrite Eq.\,(\ref{eq114}) in terms of the vapour generation rate;
\begin{eqnarray}
\frac{\dot{V}_\mathrm{vap}}{V}\approx\frac{S_{\!V}}{gh}
\frac{4u_\ast^3}{\varkappa}\ln\frac{Hu_\ast}{2\xi_0\nu}
=\frac{4S_{\!V}}{gh}u_\ast^2\sqrt{\frac{2(\delta{W_\mathrm{liq,k}}/\delta{V})}{\rho}}
\nonumber\\[5pt]
=\frac{4S_{\!V}}{gh}u_\ast^2\sqrt{\frac{\Delta\rho gh+2(\sigma_1+\sigma_2)S_{\!V}}{\rho}}\,,\quad
\label{eq119}
\end{eqnarray}
where we have employed relations (\ref{eq108}) and (\ref{eq113}) for $W_\mathrm{liq,k}$. From Eqs.\,(\ref{eq108}) and (\ref{eq113}),
\begin{equation}
\Delta\rho g\frac{h}{2}+(\sigma_1+\sigma_2)S_{\!V}\approx
\frac{\rho u_\ast^2}{2\varkappa^2}\ln^2\frac{u_\ast}{e\xi_0\nu S_{\!V}}\,.
\label{eq120}
\end{equation}
Eqs.\,(\ref{eq119})--(\ref{eq120}) determine relation between $S_{\!V}$ and $(\dot{V}_\mathrm{vap}/V)$ with $u_\ast$ as a parameter of this relation.

Unfortunately, the analytical calculation of an explicit dependence of $S_{\!V}$ on the governing parameter $(\dot{V}_\mathrm{vap}/V)$ from Eqs.\,(\ref{eq119})--(\ref{eq120}) is problematic even though the argument of the logarithm function is a large number. However, at the end of the previous section the system has been shown to be surface-tension dominated and one can typically neglect the gravitational potential energy against the background of a surface tension energy. Then Eqs.\,(\ref{eq119})--(\ref{eq120}) read
\begin{equation}
\frac{\dot{V}_\mathrm{vap}}{V}\approx
 \frac{4S_{\!V}}{gh}u_\ast^2\sqrt{\frac{2(\sigma_1+\sigma_2)S_{\!V}}{\rho}}\,,
\label{eq121}
\end{equation}
\begin{equation}
(\sigma_1+\sigma_2)S_{\!V}\approx
\frac{\rho u_\ast^2}{2\varkappa^2}\ln^2\frac{u_\ast}{e\xi_0\nu S_{\!V}}\,.
\label{eq122}
\end{equation}
From Eq.\,(\ref{eq121}),
\[
u_\ast^2\approx
\frac{gh}{4}\sqrt{\frac{\rho}{2(\sigma_1+\sigma_2)}}\frac{(\dot{V}_\mathrm{vap}/V)}{S_{\!V}^{3/2}}\,;
\]
therefore, Eq.\,(\ref{eq122}) can be rewritten as
\begin{eqnarray}
(\sigma_1+\sigma_2)S_{\!V}\approx
 \frac{\rho gh}{8\varkappa^2} \sqrt{\frac{\rho}{2(\sigma_1+\sigma_2)}}\frac{(\dot{V}_\mathrm{vap}/V)}{S_{\!V}^{3/2}}\qquad\quad
\nonumber\\[5pt]
 \times\ln^2\left(\frac{\sqrt{gh}}{2e\xi_0\nu}
 \left(\frac{\rho}{2(\sigma_1+\sigma_2)}\right)^{1/4}
 \frac{(\dot{V}_\mathrm{vap}/V)^{1/2}}{S_{\!V}^{7/4}}\right)\,.
\label{eq123}
\end{eqnarray}
Taking a square root of the later equation, one can recast it in the form
\begin{equation}
S_{\!V}^{5/4}=\alpha_1(\dot{V}_\mathrm{vap}/V)^{1/2}
 \ln{\frac{\alpha_2(\dot{V}_\mathrm{vap}/V)^{1/2}}{S_{\!V}^{7/4}}}\,,
\label{eq124}
\end{equation}
where
\begin{equation}
\alpha_1\equiv\frac{\sqrt{gh}}{2\varkappa}\left(\frac{\rho}{2(\sigma_1+\sigma_2)}\right)^{3/4},
\label{eq125}
\end{equation}
\begin{equation}
\alpha_2\equiv\frac{\sqrt{gh}}{2e\xi_0\nu}
 \left(\frac{\rho}{2(\sigma_1+\sigma_2)}\right)^{1/4}.
\label{eq126}
\end{equation}
The dependence of $S_{\!V}$ on $(\dot{V}_\mathrm{vap}/V)$ has to be derived from Eq.~(\ref{eq124}).

\begin{figure*}[!t]
\centerline{
{\sf (a)}\hspace{-5mm}
\resizebox{0.410\columnwidth}{!}{
\includegraphics%[width=0.3735\textwidth]%
 {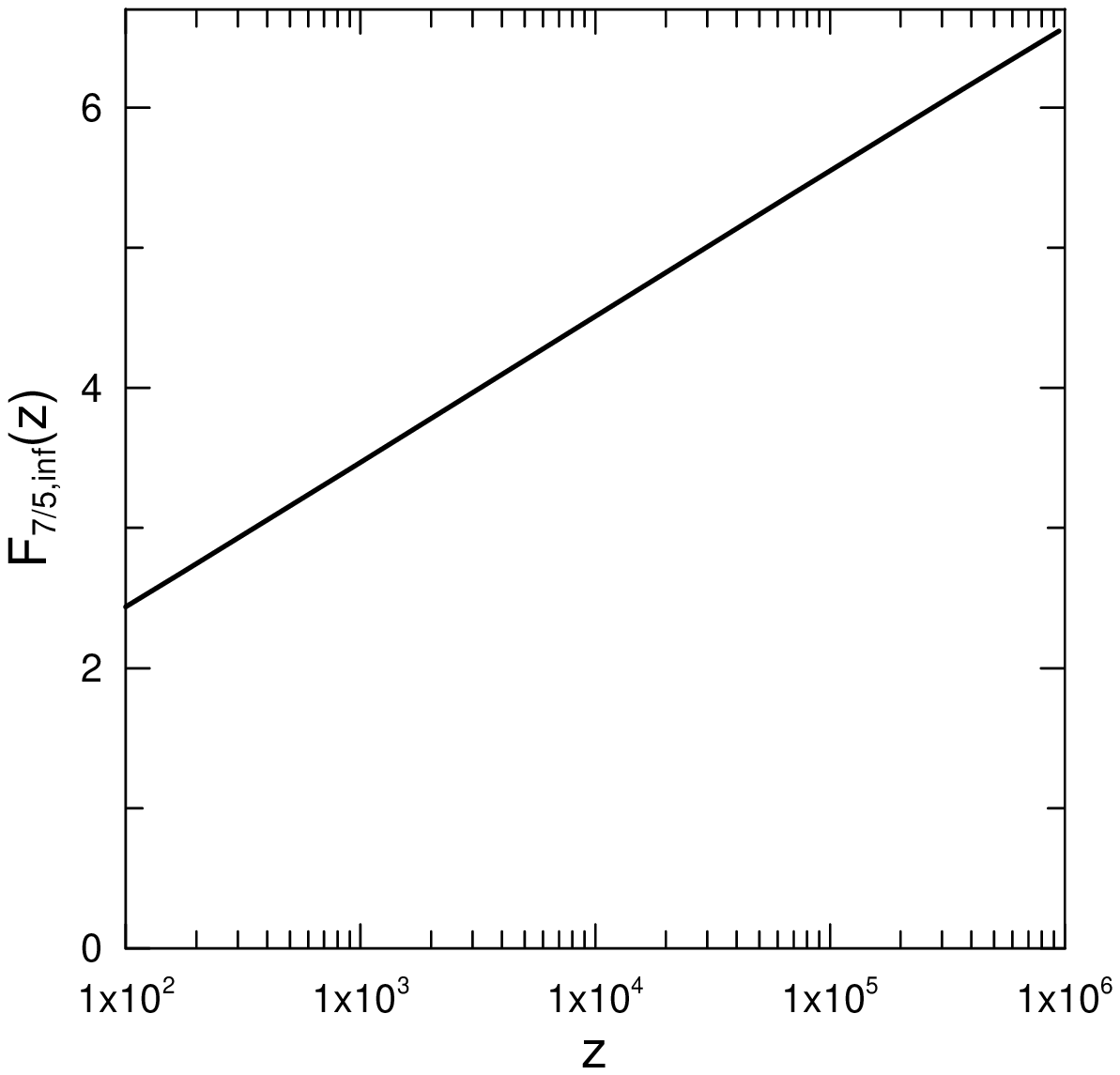}}
\qquad\quad
{\sf (b)}\hspace{-5mm}
\resizebox{0.440\columnwidth}{!}{
\includegraphics%[width=0.399\textwidth]%
 {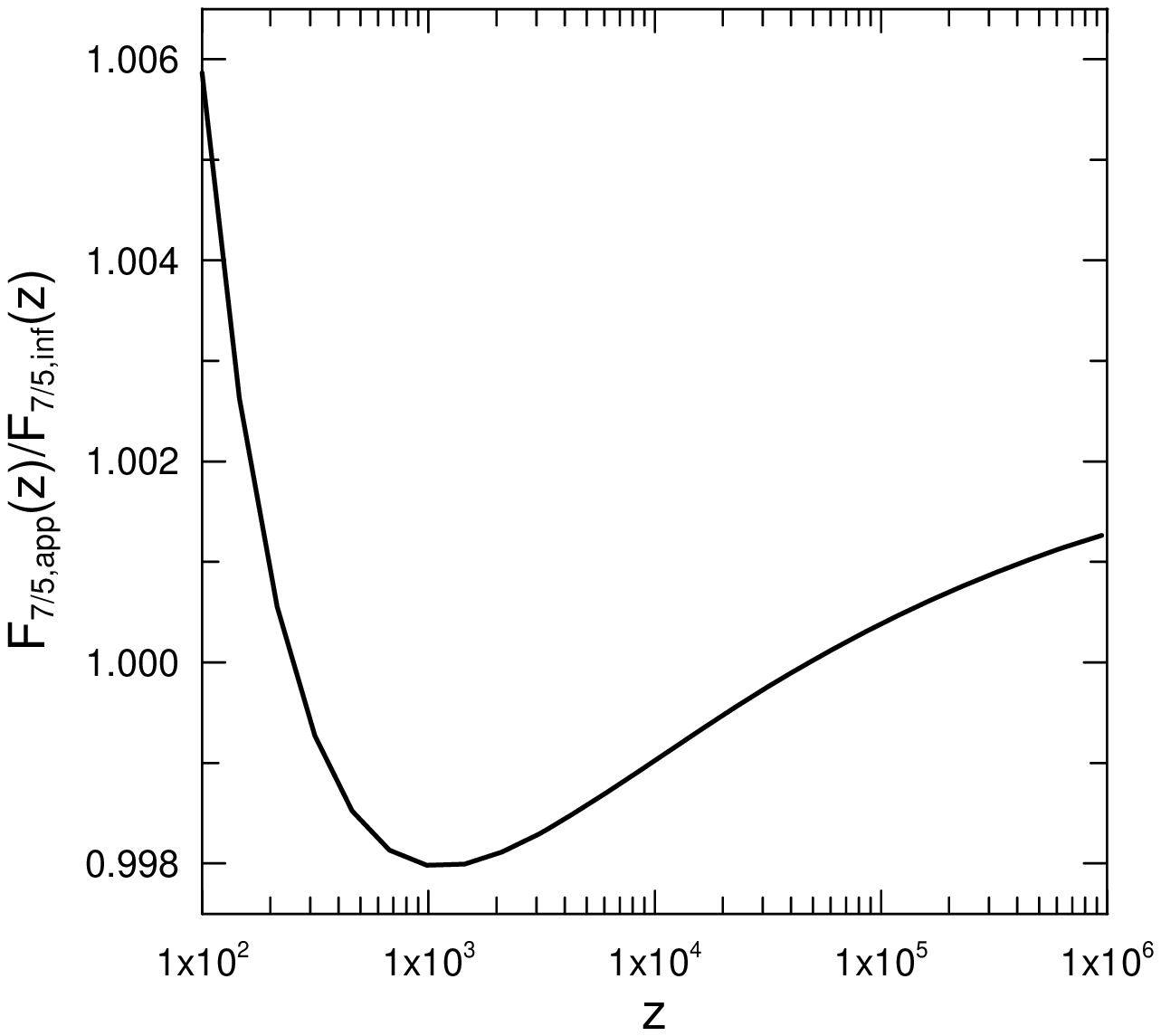}}
}

  \caption{
(a): function $F_{7/5,\infty}(z)$, (b): the ratio of approximate and exact functions $F_{7/5,\mathrm{app}}(z)/F_{7/5,\infty}(z)$
 }
  \label{fig3}
\end{figure*}

Notice, the argument of the logarithm function in Eq.~(\ref{eq124}) is of the same order of magnitude as in Eq.~(\ref{eq107}), which can be estimated for the $n$-heptane--water system with material parameters from~\cite{Pimenova-Goldobin-2014-2} and turns out to be not smaller than $10^2$--$10^3$. One can exploit this to solve Eq.~(\ref{eq124}) iteratively. Indeed, multiplying a large argument of the logarithm function by factor which can be non-small compared to $1$ but small compared to the argument, one changes the logarithm only slightly. At the $0$-th iteration,
\[
(S_{\!V}^{(0)})^{5/4}=\alpha_1(\dot{V}_\mathrm{vap}/V)^{1/2},
\]
and
\[
S_{\!V}^{(0)}=\alpha_1^{4/5}(\dot{V}_\mathrm{vap}/V)^{2/5}.
\]
At the $1$-st iteration,
\[
(S_{\!V}^{(1)})^{5/4}=\alpha_1(\dot{V}_\mathrm{vap}/V)^{1/2}
 \ln{\frac{\alpha_2\,(\dot{V}_\mathrm{vap}/V)^{1/2}}{(S_{\!V}^{(0)})^{7/4}}}\,,
\]
and
\[
S_{\!V}^{(1)}=\alpha_1^{4/5}(\dot{V}_\mathrm{vap}/V)^{2/5}
 \ln^{4/5}{\frac{\alpha_2}{\alpha_1^{7/5}\,(\dot{V}_\mathrm{vap}/V)^{1/5}}}\,.
\]
At the $2$-nd iteration,
\[
(S_{\!V}^{(2)})^{5/4}=\alpha_1(\dot{V}_\mathrm{vap}/V)^{1/2}
 \ln{\frac{\alpha_2\,(\dot{V}_\mathrm{vap}/V)^{1/2}}{(S_{\!V}^{(1)})^{7/4}}}\,,
\]
and
\begin{equation}
S_{\!V}^{(2)}=\alpha_1^\frac{4}{5}(\dot{V}_\mathrm{vap}/V)^\frac{2}{5}
 \ln^\frac{4}{5}\left(\frac{\alpha_{\dot{Q}_V}}{(\dot{V}_\mathrm{vap}/V)^\frac{1}{5}}
 \ln^{-\frac{7}{5}}{\frac{\alpha_{\dot{Q}_V}}{(\dot{V}_\mathrm{vap}/V)^\frac{1}{5}}}
 \right)\,,
\label{eq127}
\end{equation}
where
\[
\alpha_{\dot{Q}_V}\equiv\frac{\alpha_2}{\alpha_1^{7/5}}\,.
\]
Hence,
\begin{eqnarray}
S_{\!V}=\alpha_1^\frac{4}{5}(\dot{V}_\mathrm{vap}/V)^\frac{2}{5}
 \ln^\frac{4}{5}\left(\frac{\alpha_{\dot{Q}_V}}{(\dot{V}_\mathrm{vap}/V)^\frac{1}{5}}
 \ln^{-\frac{7}{5}}\left(\frac{\alpha_{\dot{Q}_V}}{(\dot{V}_\mathrm{vap}/V)^\frac{1}{5}}
\right.\right.
\nonumber\\[5pt]
\times\left.\left.
 \ln^{-\frac{7}{5}}\left(\frac{\alpha_{\dot{Q}_V}}{(\dot{V}_\mathrm{vap}/V)^\frac{1}{5}}
 \ln^{-\frac{7}{5}}\left(\frac{\alpha_{\dot{Q}_V}}{(\dot{V}_\mathrm{vap}/V)^\frac{1}{5}}
 \Bigg(\dots\Bigg)
 \right)\right)\right)\right)
\nonumber\\[5pt]
\equiv\alpha_1^\frac{4}{5}(\dot{V}_\mathrm{vap}/V)^\frac{2}{5}
F_{\frac75,\infty}\left(\frac{\alpha_{\dot{Q}_V}}{(\dot{V}_\mathrm{vap}/V)^\frac{1}{5}}\right)\,,
\qquad
\label{eq128}
\end{eqnarray}
where
\[
F_{\frac75,n}(z)
 \equiv\ln^\frac45\Big( \underbrace{z\ln^{-\frac75}\Big( \dots z\ln^{-\frac75}\Big(}_{\mbox{\footnotesize ($n-1$) times}}
 z\, \Big) \dots \Big)\Big)
\]
for $n=1,2,3,...$\,. Function $F_{7/5,\infty}(z)$ is plotted in Fig.~\ref{fig3}(a).

For different tasks, one may employ Eq.~(\ref{eq128}) to calculate $S_{\!V}$ for either given $(\dot{V}_\mathrm{vap}/V)$ or given heat inflow $\dot{Q}$. For the latter case, one has to recall the relation (\ref{eq111}) between the vapour production and heat inflow,
\[
\frac{\dot{V}_\mathrm{vap}}{V}=\frac{\dot{Q}_V}
 {\Lambda_1n_1^{(0)}+\Lambda_2n_2^{(0)}}\,.
\]

%%%%%%%%%%%%%%%%%%%%%%%%%%%%%%%%%%%%%%%%%%%%%%%%%%%%%%%%%
%%%%%%%%%%%%%%%%%%%%%%%%%%%%%%%%%%%%%%%%%%%%%%%%%%%%%%%%%

\subsection{Self-cooling}
Let us consider the dynamics of cooling-down of a system experiencing the interfacial boiling without external heat supply. The heat consumed for the vapour generation is provided from decrease of the mean system temperature;
\begin{equation}
\dot{Q}_V=\rho c_P(-\langle\dot{\varTheta}\rangle)\,.
\label{eq129}
\end{equation}
Eq.~(\ref{eq129}) expresses how the temperature decrease rate controls the heat inflow $\dot{Q}_V$ to the interface and thus the vapour production rate $\dot{V}_\mathrm{vap}$ (see Eq.~(\ref{eq111})). Then Eq.~(\ref{eq119}) yields
\begin{equation}
\frac{\rho c_P(-\langle\dot{\varTheta}\rangle)}{\Lambda_1n_1^{(0)}+\Lambda_2n_2^{(0)}}
=\frac{4S_{\!V}}{gh}u_\ast^2\sqrt{\frac{\Delta\rho gh+2(\sigma_1+\sigma_2)S_{\!V}}{\rho}}\,,\quad
\label{eq130}
\end{equation}
Eqs.~(\ref{eq130}) and (\ref{eq120}) form a self-contained equation system which determines the relation between $S_{\!V}$ and $(-\langle\dot{\varTheta}\rangle)$ with $u_\ast$ as a parameter of this relation.

%%%%%%%%%%%%%%%%%%%%%%%%%%%%%%%%%%%%%%%%%%%%%%%%%%%%%%%%%
\begin{figure*}[!t]
\centerline{
{\sf (a)}\hspace{-8mm}
\resizebox{0.431\columnwidth}{!}{
\includegraphics%[width=0.39\textwidth]%
 {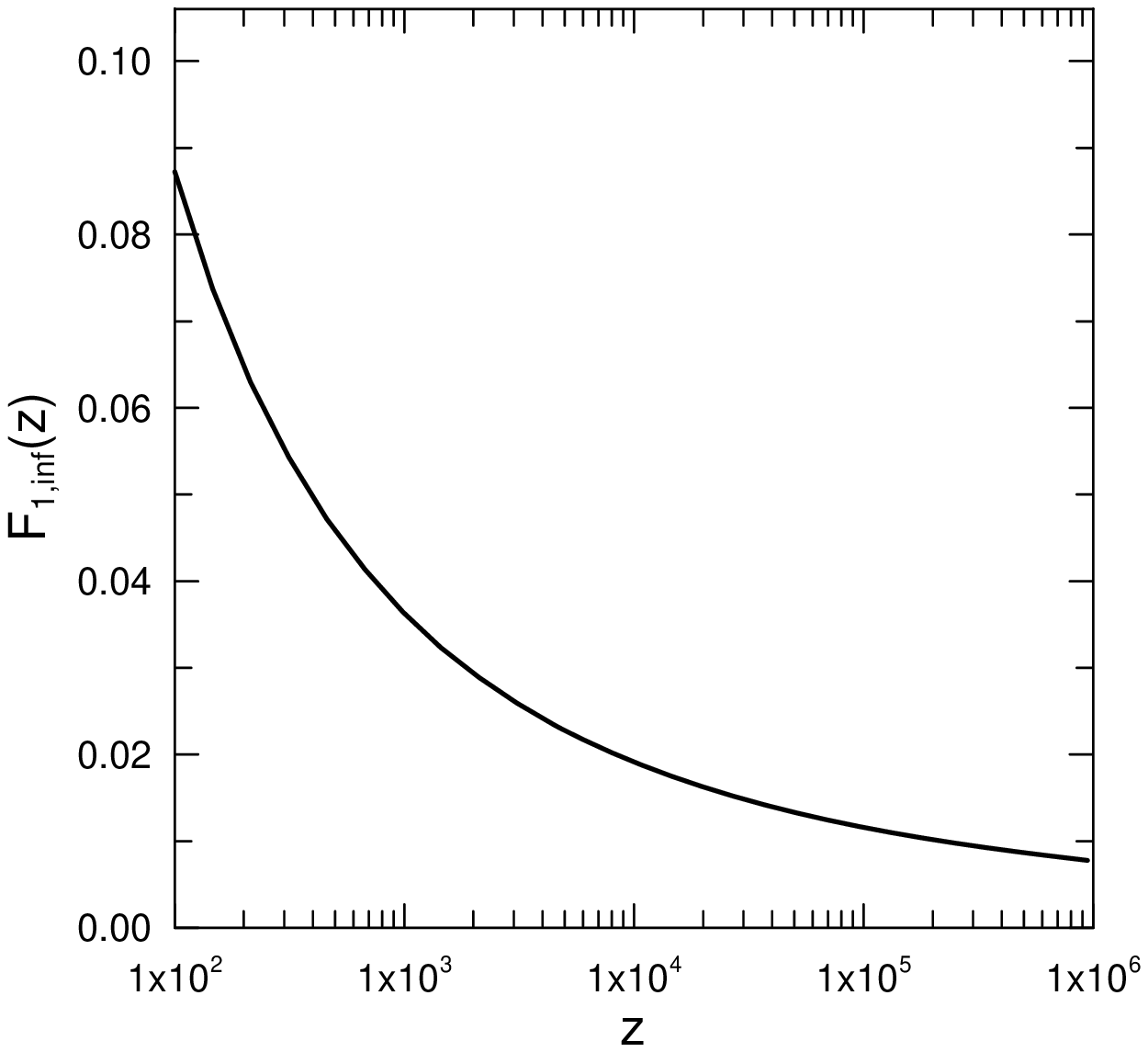}}
\qquad\quad
{\sf (b)}\hspace{-5mm}
\resizebox{0.440\columnwidth}{!}{
\includegraphics%[width=0.399\textwidth]%
 {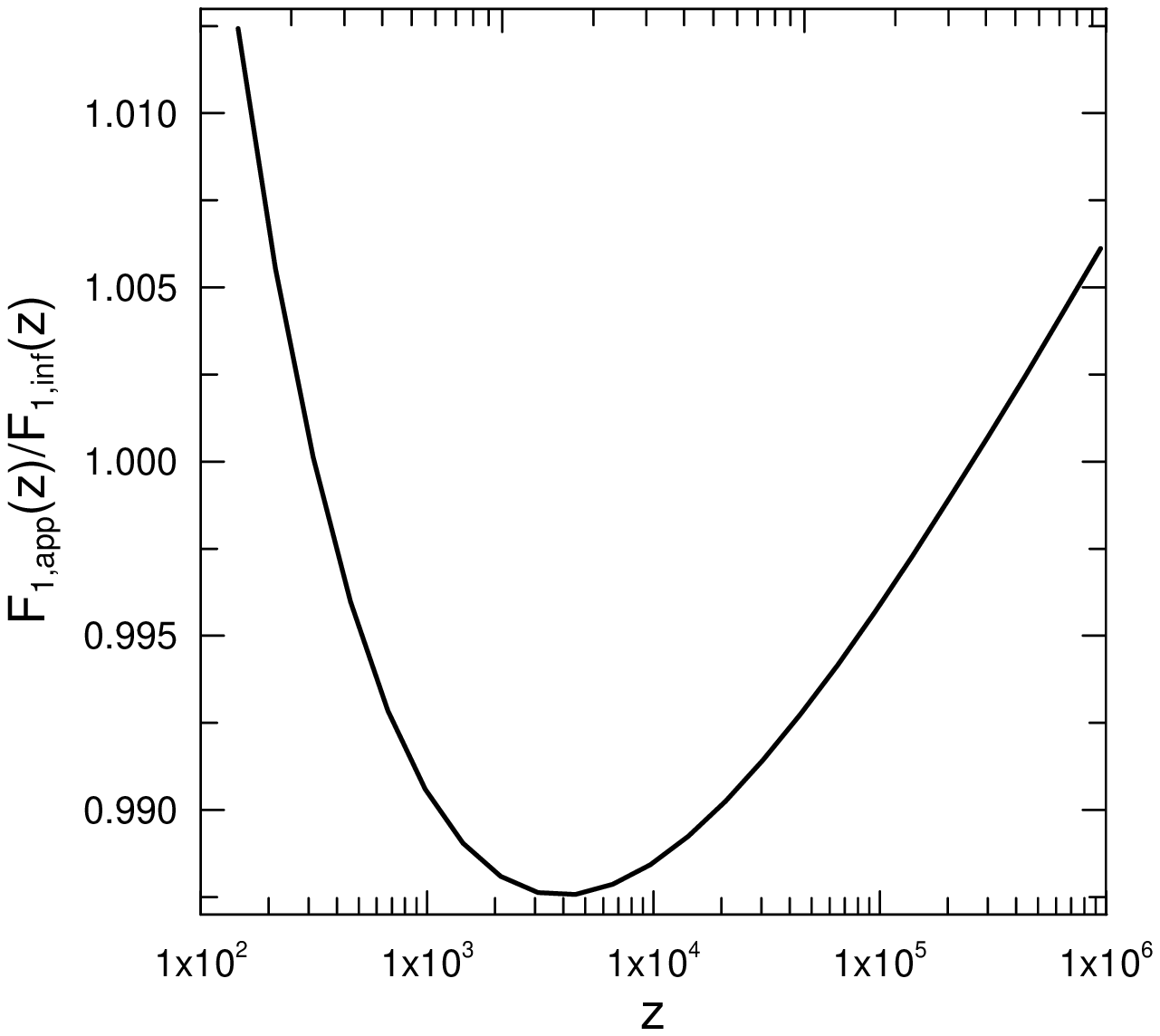}}
}

  \caption{
(a): function $F_{1,\infty}(z)$, (b): the ratio of approximate and exact functions  $F_{1,\mathrm{app}}(z)/F_{1,\infty}(z)$
 }
  \label{fig4}
\end{figure*}
%%%%%%%%%%%%%%%%%%%%%%%%%%%%%%%%%%%%%%%%%%%%%%%%%%%%%%%%%

Until the latest stages of self-cooling the mean temperature excess $\langle\varTheta\rangle$ stays large compared to $\varTheta_g$, and one can consider the system to be surface-tension dominated. Then, similarly to Eqs.~(\ref{eq121})--(\ref{eq122}), one finds from Eqs.~(\ref{eq130}) and (\ref{eq120}):
\begin{equation}
\frac{\rho c_P(-\langle\dot{\varTheta}\rangle)}{\Lambda_1n_1^{(0)}+\Lambda_2n_2^{(0)}}
\approx
 \frac{4}{gh}\sqrt{\frac{2(\sigma_1+\sigma_2)}{\rho}}u_\ast^2S_{\!V}^{3/2}\,,
\label{eq131}
\end{equation}
\begin{equation}
(\sigma_1+\sigma_2)S_{\!V}\approx
\frac{\rho u_\ast^2}{2\varkappa^2}\ln^2\frac{u_\ast}{e\xi_0\nu S_{\!V}}\,.
\label{eq132}
\end{equation}

From Eq.\,(\ref{eq131}),
\[
u_\ast^2\approx
\frac{c_P(-\langle\dot{\varTheta}\rangle)\,gh}{4(\Lambda_1n_1^{(0)}+\Lambda_2n_2^{(0)})}
 \frac{1}{\sqrt{2(\sigma_1+\sigma_2)}}\left(\frac{\rho}{S_{\!V}}\right)^{3/2}\,;
\]
and Eq.\,(\ref{eq132}) yields
\begin{equation}
S_{\!V}^{5/4}=\widetilde{\alpha}_1{\textstyle\sqrt{-\langle\dot\varTheta\rangle}}
 \ln{\frac{\widetilde{\alpha}_2\sqrt{-\langle\dot\varTheta\rangle}}{S_{\!V}^{7/4}}}\,,
\label{eq133}
\end{equation}
where
\begin{equation}
\widetilde{\alpha}_1\equiv\frac{1}{2\varkappa}
\sqrt{\frac{\rho c_P\,gh}{\Lambda_1n_1^{(0)}+\Lambda_2n_2^{(0)}}}
 \left(\frac{\rho}{2(\sigma_1+\sigma_2)}\right)^{3/4},
\label{eq134}
\end{equation}
\begin{equation}
\widetilde{\alpha}_2\equiv\frac{1}{2e\xi_0\nu}
 \sqrt{\frac{\rho c_P\,gh}{\Lambda_1n_1^{(0)}+\Lambda_2n_2^{(0)}}}
 \left(\frac{\rho}{2(\sigma_1+\sigma_2)}\right)^{1/4}.
\label{eq135}
\end{equation}

Employing relation (\ref{eq118}), one can find
\[
\dot{S}_V=\frac{k_{12}^2h}{2\varTheta_g}\langle\dot\varTheta\rangle
\]
and recast Eq.~(\ref{eq135}) in the forms
\begin{equation}
S_{\!V}^{5/4}=\gamma_1{\textstyle\sqrt{-\dot{S}_{\!V}}}
 \ln{\left(\gamma_2\frac{\sqrt{-\dot{S}_{\!V}}}{S_{\!V}^{7/4}}\right)}\,,
\label{eq136}
\end{equation}
where
\begin{equation}
\gamma_{1,2}\equiv\frac{2\varTheta_g}{k_{12}^2h}\widetilde{\alpha}_{1,2}\,.
\label{eq137}
\end{equation}

One can transform Eq.~(\ref{eq136}) into the equation of dynamics similarly to the transformation of Eq.~(\ref{eq124}) into Eq.~(\ref{eq128});
\[
-\dot{S}_{\!V}=\frac{S_{\!V}^{5/2}}{\gamma_1^2}F_{1,\infty}
\left(\frac{\gamma_2/\gamma_1}{S_{\!V}^{1/2}}\right)\,,
\]
where
\[
F_{1,n}(z)
 \equiv\ln^{-2}\Big( \underbrace{z\ln^{-1}\Big( \dots z\ln^{-1}\Big(}_{\mbox{\footnotesize ($n-1$) times}}
 z\, \Big) \dots \Big)\Big)\,.
\]
The latter continuous fraction converges well when the argument $z$ is large.
In dimensionless terms, one can write
\begin{equation}
\frac{\mathrm{d}\eta}{\mathrm{d}\tau}=F_{1,\infty}\left(\eta^2\right)\,,
\label{eq138}
\end{equation}
where dimensionless time
\[
\tau=\frac{t}{4\gamma_1^2}\sqrt{\frac{\gamma_2}{\gamma_1}}
\]
and
\[
\eta=\frac{1}{S_{\!V}^{1/4}}\sqrt{\frac{\gamma_2}{\gamma_1}}\,.
\]

\subsection{Approximation for $F_{7/5,\inf}(z)$ and $F_{1,\inf}(z)$}
Functions $F_{7/5,\inf}(z)$ and $F_{1,\inf}(z)$ can be approximated with
\begin{equation}
F_{7/5,\mathrm{app}}(z)=\left(1.083+(F_{7/5,2}(z))^{1.64}\right)^{1/1.64}\,,
\label{eq139}
\end{equation}
\begin{equation}
F_{1,\mathrm{app}}(z)=0.745943(F_{1,2}(z))^{0.9451}\,,
\label{eq140}
\end{equation}
respectively. In Figures~\ref{fig3} and \ref{fig4}, one can see that relative error of these approximations in the relevant range of values of $z$ is about $1\%$.

\section{Conclusion}
For the system of two immiscible liquids experiencing interfacial boiling we have assessed the value of the mean specific interface area $S_{\!V}\equiv(\delta{S}/\delta{V})$ as a function of macroscopic characteristics of the system state:
\\
$\bullet$\ the mean overheat $\langle\varTheta\rangle$ above the minimal temperature of interfacial boiling $T_\ast$ (Eq.~(\ref{eq118})),
\\
$\bullet$\ the vapour volume production per the unit volume of the system $(\dot{V}_\mathrm{vap}/V)$ (Eq.~(\ref{eq128})), {\it or}
\\
$\bullet$\ the heat inflow to the system $\dot{Q}_V$ (Eqs.~(\ref{eq128}) and (\ref{eq111})).
\\
The calculations are based on the mechanical energy flux balance in the system and the assumption of system stochatization. The system is assumed to be well-stirred by boiling. The heat and momentum transfer towards the contact interface is considered to be turbulent one and obeys the theory of the turbulent boundary layer.

With the dependencies between macroscopic characteristics, which we have obtained in this paper, one can construct a comprehensive self-contained mathematical model of the process of interfacial boiling~\cite{Pimenova-Goldobin-JETP-2014,Pimenova-Goldobin-2014-2}. This model will be valid for the case of the system of two immiscible liquids below the bulk boiling temperature of both components.

For the case of no heat supply, within the framework of this approach, we have derived the equation of self-cooling dynamics of the system (\ref{eq138}).

The results are provided in the form of continued fraction--logarithm, which possess slow convergence properties even though the arguments are as large as $10^2$--$10^3$. For faster calculations we can suggest the approximations (\ref{eq139}) and (\ref{eq140}), the relative error of which is about $1\%$.

The work has been financially supported by the Russian Science Foundation (grant no.\ 14-21-00090).

\end{document}